\documentclass[sigconf]{acmart}

\usepackage{xargs}

\usepackage{algorithm}
\usepackage[noend]{algpseudocode}

\newcommand{\theproduct}{\textit{SmartKex\xspace}}

\usepackage{listings}
\usepackage{xcolor}
\usepackage{multirow}
\usepackage{colortbl}
\usepackage{hhline}
\usepackage{xspace}
\usepackage[frozencache=true,cachedir=minted-cache]{minted} 
\usepackage{subcaption}

\AtBeginDocument{%
  \providecommand\BibTeX{{%
    \normalfont B\kern-0.5em{\scshape i\kern-0.25em b}\kern-0.8em\TeX}}}

\setcopyright{acmcopyright}
\copyrightyear{2022}
\acmYear{2022}

\acmConference[ ]{}{}

\renewcommand\footnotetextcopyrightpermission[1]{} 

\begin{document}

\title{\theproduct: Machine Learning Assisted SSH Keys Extraction From The Heap Dump}

\author{Christofer Fellicious}
\authornote{Both authors contributed equally to this research.}
\email{christofer.fellicious@uni-passau.de}
\author{Stewart Sentanoe}
\authornotemark[1]
\email{se@sec.uni-passau.de}
\affiliation{%
  \institution{University of Passau}
  \streetaddress{Innstr. 41}
  \city{Passau}
  \state{Bayern}
  \country{Germany}
  \postcode{94032}
}

\author{Michael Granitzer}
\email{michael.granitzer@uni-passau.de}
\affiliation{%
  \institution{University of Passau}
  \streetaddress{Innstr. 41}
  \city{Passau}
  \state{Bayern}
  \country{Germany}
  \postcode{94032}
}

\author{Hans P. Reiser}
\email{hansr@ru.is}
\affiliation{%
  \institution{Reykjavík University}
  \city{Reykjavík}
  \country{Iceland}
}
\affiliation{%
  \institution{University of Passau}
  \city{Passau}
  \state{Bayern}
  \country{Germany}
}

\renewcommand{\shortauthors}{Fellicious and Sentanoe, et al.}

\begin{abstract}
Digital forensics is the process of extracting, preserving, and documenting evidence in digital devices. 
A commonly used method in digital forensics is to extract data from the main memory of a digital device. However, the main challenge is identifying the important data to be extracted.
Several pieces of crucial information reside in the main memory, like usernames, passwords, and cryptographic keys such as SSH session keys.
In this paper, we propose~\theproduct, a machine-learning assisted method to extract session keys from heap memory snapshots of an OpenSSH process. In addition, we release an openly available dataset and the corresponding toolchain for creating additional data.
Finally, we compare~\theproduct~ with naive brute-force methods and empirically show that \theproduct~ can extract the session keys with high accuracy and high throughput. With the provided resources, we intend to strengthen the research on the intersection between digital forensics, cybersecurity, and machine learning. 

\end{abstract}

\keywords{datasets, memory dump, secure-shell, session keys extraction, machine learning, random forest, supervised learning}


\maketitle

\section{Introduction}\label{sec:introduction}

Digital forensics plays a vital role in the digital era.
It helps us extract evidence from modern devices such as mobile phones, laptops, PCs, and so forth. 
This evidence can help to detect malicious software or digital fingerprints of intruders.
A primary method to extract evidence analyses the main memory contents from a device.
However, without any knowledge about the extracted data, the process becomes too complex to obtain any useful information~\cite{jain2014sok}.

Here, we leverage machine learning methods to help the investigator to extract valid information from the main memory.
SSH is a secure method to communicate with a remote server.
Its security is not only beneficial for benign users, it also helps malicious attackers to keep their actions secret.
Decrypting potentially malicious SSH connections can thus be a very valuable means in a forensic investigation.
Our research focuses on extracting session keys of the SSH protocol from the memory heap.
On that basis, our research will help the analyst to investigate malicious activities on a remote server by decrypting the network traffic.

Machine learning has applications in various domains, from image classification~\cite{krizhevsky2012imagenet} to music generation~\cite{yang2017midinet}. 
Recently, more and more machine learning applications have also focused
on security, mainly detecting malware from process information. 
Gibert et al. give an overview of implementing
machine learning methods into dynamic analysis tools to improve malware detection.~\cite{gibert2020rise}. However, machine learning on raw main memory is often underutilized to the best of our knowledge, primarily due to the potentially high data volumes to be processed.

In this work, we contribute to machine learning techniques for main memory analysis in digital forensics through 
\begin{enumerate}
    \item an open, annotated dataset of the heap memory obtained from different OpenSSH versions with different scenarios. In total, it contains more than 90,000 training samples and more than 16,000 validation samples.
    \item a tool to extract session keys of OpenSSH from the main memory;
    \item a naive brute-force method to extract SSH keys from the heap dump;
    \item a machine learning assisted method for efficiently extracting SSH Keys from the heap dump.
\end{enumerate}
Using the combination of the brute force method and the random forest classifier, we show that our method succeeds in reducing the search area by more than 90\%. Moreover, it takes less than 1$s$ to extract the session keys. 


\section{Background}\label{sec:background}
This section gives a bit of background information about the technologies related to this paper. 

\subsection{Secure Shell (SSH)}
Secure Shell (SSH) is a network protocol that allows a user to communicate securely with a remote resource over an insecure network.
To ensure confidentiality, each SSH session uses a set of session keys comprised of six keys:
\begin{itemize}
    \item Key A: Initialization vector (IV) from client to server
    \item Key B: Initialization vector (IV) from server to client
    \item Key C: Encryption key (EK) from client to server
    \item Key D: Encryption key (EK) from server to client
    \item Key E: Integrity key from client to server
    \item Key F: Integrity key from server to client
\end{itemize}

It is necessary to know about the IV and EK pair (Key A and C, or Key B and D) to decrypt an encrypted SSH session's network traffic (assuming that the passive network monitoring is exists).

The most commonly used SSH implementation is OpenSSH~\cite{openssh}. For this paper, we are using OpenSSH from V6\_0P1 until V8\_8P1. OpenSSH offers several encryption methods, such as Advanced Encryption Standard (AES) Cipher Block Chaining (CBC)\cite{frankel2003aes}, AES Counter (AES-CTR)\cite{lipmaa2000comments}, and ChaCha20\cite{bernstein2008chacha}. Each encryption method also has different IV and EK key lengths between 12 and 64 bytes.

\subsection{Heap Memory}
Heap memory (also dynamic memory) is an alternative to local stack memory. Local stack memory stores local variables when a function is called
and deallocates them at the function exit. On the other hand, heap requires the program to explicitly request the allocation of memory (using \textit{new} operator in Java and C++, or \textit{malloc}/\textit{calloc} in C). 
The allocated memory can be deallocated automatically via the garbage collector (Java) or by manually deallocating them using the \textit{delete} or \textit{free} operator (C and C++).

In our case, OpenSSH is written in C and uses \textit{calloc} to allocate memory blocks that hold the session data, including the keys.
Using this information, if we dump the heap of a running OpenSSH process at the correct time (e.g. in the middle of an active SSH session), the dumped heap file will contain the SSH session keys.

\section{Related Work}\label{sec:related-work}

Purnaye and Kulkarni created memory dumps of virtual machines containing around 360 Virtual Machine (VM) dumps with a total dataset file size of approximately 80GB zipped\cite{prasad2020}. This dataset contains dumps of continuously generated data. 
The Dumpware10 dataset covers 4294 samples from 10 different malware families~\cite{bozkir2021catch}. This dataset contains 3686 malware and 608 benign samples.
The dataset can be represented as RGB images and has the advantage that computer vision methods are compatible with the dataset. Sadek et al. created a dataset compromising Windows 10 VMs~\cite{sadek2019memory}. The authors deliberately infected Windows 10 VMs and collected ten snapshots of the VM once the malicious payload was running. The snapshots are in the "Advanced Forensic Format" (AFF4), which is a compressed format. The compressed file size is around 1GB per snapshot. The dataset contains 1530 snapshots.

Petrik et al. developed a method to analyze raw binary data extracted from the memory dump of a device~\cite{petrik2018towards}. The authors use machine learning methods and a multi-hundred Terabyte dataset to detect malware in memory dumps with a very high success rate. This model aims to be architecture and Operating System (OS) independent for malware detection. Sihwail et al. used a method that combined memory forensics to extract malicious artefacts and generate features for machine learning. The authors report a very high accuracy and low false positive rate. Tran et al. also use memory forensics and machine learning to identify malware data~\cite{tran2021independent}. They implement an OS-independent malware detector which is also geared towards finding unidentified malware.

\section{Dataset Generation}\label{sec:dataset}
We use \textit{SSHKex}~\cite{sentanoe2022sshkex} as the primary method to extract the SSH keys from the main memory.
In addition, we add two features to \textit{SSHKex}: automatically dump OpenSSH's heap and add support for SSH client monitoring.

For this paper, we are using four SSH scenarios: the client connects to the server and exits immediately, port-forward, secure copy, and SSH shared connection~\cite{openssh-shared-conn}.


Two file formats, JSON and RAW, are used to store the generated logs. The JSON log file contains meta information such as the encryption name, the virtual memory address of a key, and the key's value in hex representation (as shown in Figure~\ref{fig:json_log}). The binary file contains the heap dump of the OpenSSH process (as shown in Figure~\ref{fig:heap_dump} using the \textit{xxd} command). 



\begin{figure}[tb]
    \centering
\begin{minted}[frame=single, rulecolor=black, fontsize=\footnotesize, linenos=false, style=murphy,
label=\textit{Part of OpenSSH session keys' log}]{json}
{
    ...
    "ENCRYPTION_KEY_1_NAME": "aes192-ctr",
    "KEY_A_ADDR": "55a5c37449a0",
    "KEY_A_LEN": 16,
    "KEY_A": "564084fff3c69eed22e9c59b7d46b6d0",
    ...
}
\end{minted}
\caption{Sample of an OpenSSH key extraction log file}
\label{fig:json_log}
\end{figure}

\begin{figure}[tb]
    \centering
\begin{minted}[frame=single, rulecolor=black, fontsize=\footnotesize, linenos=false, style=murphy,
label=\textit{Part of OpenSSH's heap}]{text}
000225c0: 0000 0000 0000 0000 2100 0000 0000 0000  ........!.......
000225d0: 6165 7331 3932 2d63 7472 0000 0000 0000  aes192-ctr......
000225e0: 2000 0000 0000 0000 9100 0000 0000 0000   ...............
000225f0: 4015 74c3 a555 0000 10f0 71c3 a555 0000  @.t..U....q..U..
\end{minted}
\caption{Sample of an OpenSSH heap dump}
\label{fig:heap_dump}
\end{figure}

There are two top-level directories in the dataset: training and validation. Each of these top-level directories are further divided into subdirectories based on the scenario used such as OpenSSH, port-forwarding or secure copy (SCP). 

The subdirectories below the OpenSSH or SCP are split based on the software version that created the memory dump.
These directories are again organized into different directories based on the software version that created the memory dump. 
We further organize the heaps based on their key lengths, with each key length having its directory below the version directory.
These version directories are further divided into the different key lengths in a heap. 
A JSON file with the same alphanumeric sequence except for the "-heap" part accompanies every raw memory dump. 
The JSON file contains the different encryption keys and other metadata such as the process id, the offset of the heap, and so forth\footnote{For some versions, it was only possible to extract the encryption keys}. Thus, the dataset is not limited to the task of extracting session keys but also in identifying the essential data structures that hold sensitive information. The dataset, code and tools are open-sourced. The dataset is available in a Zenodo repository\footnote{\url{https://zenodo.org/record/6537904}}. The code is available in a public Github repository\footnote{\url{https://github.com/smartvmi/Smart-and-Naive-SSH-Key-Extraction}}.

\section{Methodology}\label{sec:methodology}
This section discusses two methods to extract session keys from heap dump: a baseline and machine learning assisted methods.  The source code for all our implementations is open source and reproducible~\cite{SSHKeyEx,zenodo}.

\subsection{Baseline Method (Brute-force)}
\label{subsec:baseline}
We implement a naive brute-force method as our baseline method. To reduce the size of the heap dump, we delete memory pages that are insignificant using the hamming distance method~\cite{taubmann2016tlskex}.
Algorithm~\ref{alg:bruteforce} describes the brute-force method.

First, we set $ivLen$ and $keyLen$ based on the encryption method for the heap.
Then, we take the first $ivLen$ bytes of the heap dump as the potential IV ($pIV$).
Next, we take $keyLen$ bytes from the heap dump, starting from the first byte as the potential key ($pKey$).
After that, we iterate the potential key until it reaches the end of the heap dump. 
Finally, if decryption of the network packet is not feasible, we repeat the process by reading the subsequent potential IV and the next potential key.

We also need the network traffic recording for the brute-force method. 
A standard passive network sniffing tool like Wireshark or TCPDump records the necessary traffic for the brute force method.

\begin{algorithm}
  \caption{SSH keys brute-force algorithm}\label{alg:bruteforce}
  \begin{algorithmic}[1]
    \Procedure{FINDIVANDKEY}{$netPacket,heapDump$}
        \State $ivLen\gets 16$ \Comment{Based on the encryption method}
        \State $keyLen\gets 24$ \Comment{Based on the encryption method}
        \State $i\gets sizeof(cleanHeapDump)$
        \State $r\gets 0$
        \While{$r<i$}
            \State $pIV\gets heapDump[r:r+ivLen]$
            \State $x\gets 0$
            \While{$x<i$}
                \State $pKey\gets heapDump[x:x+keyLen]$
                \State $f\gets decrypt(netPacket, pIV, pKey)$
                
                \If{f is TRUE}
                    \State \textbf{return }$pIV,pKey$
                \EndIf
                \State $x\gets x+8$ \Comment{The IV is 8-bytes aligned}
            \EndWhile
            \State $r\gets x+8$ \Comment{The key is 8-bytes aligned}
        \EndWhile
    \EndProcedure
  \end{algorithmic}
\end{algorithm}

\subsection{Machine Learning}
\label{subsec:machine_learning}
The generated heaps' size varies between 100KB to 500KB, depending on the protocol used. As the raw heap dump can be of arbitrary shape and since many machine learning algorithms such as random forests work with fixed-sized data inputs; we preprocess and adapt the data to the machine learning model.
\subsubsection{Preprocessing}\hfill\\

Auguste Kerckhoffs' cryptographic principle states that "The principle holds that a cryptosystem should be secure, even if everything about the system, except the key, is public knowledge"~\cite{kerckhoffs1883cryptographic}. For security by openness, random numbers and their generations form the building block of encryption keys. 
Therefore, we expect encryption keys to be primarily random byte sequences and thus have a large entropy. 
By looking at 8 bytes aligned data, we can detect the high entropy parts because the encryption keys are also 8 bytes aligned in memory.
We resize the heap data into an $N\times8$ matrix, where $N*8$ will be the size of the original heap data in bytes. 
The preprocessing algorithm starts by considering the discrete differences of the bytes in the vertical and horizontal directions. We then do the logical AND operation on the horizontal and vertical absolute differences as shown in \autoref{eqn:grad_compute}. The presence of zeros means that the adjacent element vertically or horizontally has the same value. 

\begin{equation}\label{eqn:grad_compute}
\begin{aligned}
    Y[i][j] = |X[i][j] - X[i][j+1]| \& |X[i][j] - X[i+1][j]|, \\
    \text{where X is the N x 8 reshaped matrix}
\end{aligned}
\end{equation}

Each 8-byte row is then examined based on the randomness and
if half of the bytes are different from the adjacent bytes, it is possible to be part of an encryption key as given in \autoref{eqn:set_true} and marked appropriately.

\begin{equation}\label{eqn:set_true}
\begin{aligned}
    Z[i] = count(Y[i][k] == 0) >= 4, \\
    \text{where Y is the result of \autoref{eqn:grad_compute}}
\end{aligned}
\end{equation}

As the minimum length of an encryption key in our scenario is twelve bytes, there should be at least two consecutive rows marked as possible keys from \autoref{eqn:set_true}for an encryption key. 
Isolated single rows marked as possible keys are discarded due to this reason.
This removal of isolated rows is done in \autoref{eqn:find_bytes} simply by 
The result of \autoref{eqn:find_bytes} will be an array with "1"s indicating the possible locations of the keys. This operation is similar to the morphological operation "Opening" in image processing.

\begin{equation}\label{eqn:find_bytes}
\begin{aligned}
    R[i] = Z[i] \And Z[i+1], \\
    \text{where Z is a linear array from \autoref{eqn:set_true}}
\end{aligned}
\end{equation}

After discarding these bytes from \autoref{eqn:find_bytes}, we extract 128-byte sized slices (or windows) to train the model. 
With this fixed size input, we train a Random Forest algorithm with the labels 0 and 1. Label 1 if a key is present anywhere within the 128-byte slice; otherwise, the label is 0.
The additional benefit of \autoref{eqn:find_bytes} is that it lets us select specific lengths of keys by just looking at the results. If there is a need to select only blocks having encryption keys of twenty-four bytes or longer, only locations with a contiguous sequence of at least three '1' need to be checked. This encoding reduces the effort of looking at individual 8 byte aligned blocks for keys and simplifies the training. 

 

 

\begin{table*}[ht!]
   \centering
    \captionsetup[subtable]{position = below}
  \captionsetup[table]{position=top}
   \caption{Metrics for the machine learning model}
   \label{tab:ml_result}
   \begin{subtable}{0.5\textwidth}
       \centering
       \begin{tabular}{ |c|c|c|c|c| } 
         \hline
         \textbf{Classifier} & \textbf{Accuracy} & \textbf{Precision} & \textbf{Recall} & \textbf{F1-Score} \\
         \hline
         High Precision & 99.75 & 93.17 & 84.37 & 88.55 \\ 
         High Recall & 99.06 & 55.53 & 99.62 & 71.31 \\ 
         Stacked & 99.56 & 76.09 & 91.60 & 83.13 \\ 
         \hline
         
        \end{tabular}
       \caption{Metrics on labelled data}
       \label{tab:ml_result_a}
   \end{subtable}%
   \hspace*{1em}
   \begin{subtable}{0.5\textwidth}
       \centering
       \begin{tabular}{ |c|c|c|c|c| } 
         \hline
         \multicolumn{1}{|c|}{\textbf{Key Len}} & \multicolumn{1}{c|}{\textbf{Total \# Keys}} & \multicolumn{1}{c|}{\begin{tabular}[c]{@{}c@{}}\textbf{High}\\\textbf{Recall}\end{tabular}} & \multicolumn{1}{c|}{\begin{tabular}[c]{@{}c@{}}\textbf{High}\\\textbf{Precision}\end{tabular}} & \multicolumn{1}{c|}{\textbf{Stacked}} \\
         \hline
         12 & 12,422 & 12,421 & 12,331 & 12,369 \\  
         16 & 24,264 & 24,210 & 20,853 & 23,129 \\ 
         24 & 8,416 & 8,416 & 7,686 & 8,301 \\
         32 & 8,702 & 8,701 & 7,779 & 8,545 \\
         64 & 6,312 & 6,300 & 3,860 & 5,622 \\
         \hline
         
        \end{tabular}
        \caption{Key Length based Results}
         \label{tab:ml_result_b}
   \end{subtable}
\end{table*}

\begin{table*}[ht!]
\caption{Performance comparison between brute-force and machine learning methods}
\label{tab:performance-result}
\footnotesize
\centering
\begin{tabular}{|l|l|l|l|l|l|l|l|l|l|l|l|l|l|l|l|l|} 
\hline
\multicolumn{1}{|c|}{{\cellcolor{black}}} & \multicolumn{1}{c|}{\textbf{Version}} & \multicolumn{3}{c|}{\textbf{V\_7\_1\_P1}} & \multicolumn{3}{c|}{\textbf{V\_7\_8\_P1}} & \multicolumn{3}{c|}{\textbf{V\_7\_9\_P1}} & \multicolumn{3}{c|}{\textbf{V\_8\_0\_P1}} & \multicolumn{3}{c|}{\textbf{V\_8\_1\_P1}} \\ 
\hline
{\cellcolor{black}} & \multicolumn{1}{c|}{\textbf{Key len}} & \multicolumn{1}{c|}{\textbf{16}} & \multicolumn{1}{c|}{\textbf{24}} & \multicolumn{1}{c|}{\textbf{32}} & \multicolumn{1}{c|}{\textbf{16}} & \multicolumn{1}{c|}{\textbf{24}} & \multicolumn{1}{c|}{\textbf{32}} & \multicolumn{1}{c|}{\textbf{16}} & \multicolumn{1}{c|}{\textbf{24}} & \multicolumn{1}{c|}{\textbf{32}} & \multicolumn{1}{c|}{\textbf{16}} & \multicolumn{1}{c|}{\textbf{24}} & \multicolumn{1}{c|}{\textbf{32}} & \multicolumn{1}{c|}{\textbf{16}} & \multicolumn{1}{c|}{\textbf{24}} & \multicolumn{1}{c|}{\textbf{32}} \\ 
\hhline{|-----------------|}
\multirow{-2}{*}{{\cellcolor{black}}} & \multicolumn{1}{c|}{\textbf{Heap Dump (KB)}} & \multicolumn{3}{c|}{\textbf{132 (0)}} & \multicolumn{3}{c|}{\textbf{132 (0)}} & \multicolumn{3}{c|}{\textbf{264 (0)}} & \multicolumn{3}{c|}{\textbf{264 (0)}} & \multicolumn{3}{c|}{\textbf{264 (0)}} \\ 
\hhline{|=================|}
\multirow{2}{*}{\textbf{Brute-force}} & \multicolumn{1}{c|}{Clean heap (KB)} & \multicolumn{3}{c|}{32.43 (1.00)} & \multicolumn{3}{c|}{\begin{tabular}[c]{@{}c@{}}31.03 (1.22)\\\end{tabular}} & \multicolumn{3}{c|}{67.75 (3.23)} & \multicolumn{3}{c|}{68.00 (2.41)} & \multicolumn{3}{c|}{67.98 (4.46)} \\ 
\cline{2-17}
 & \begin{tabular}[c]{@{}c@{}}Average Time (s)\\Std Dev\end{tabular} & 
 \begin{tabular}[c]{@{}c@{}}7.74\\(0.59)\end{tabular} & 
 \begin{tabular}[c]{@{}c@{}}11.35\\(0.14)\end{tabular} & 
 \begin{tabular}[c]{@{}c@{}}23.01\\(0.23)\end{tabular} & 
 \begin{tabular}[c]{@{}c@{}}7.87\\(0.82)\end{tabular} & 
 \begin{tabular}[c]{@{}c@{}}11.1\\(0.66)\end{tabular} & 
 \begin{tabular}[c]{@{}c@{}}22.4\\(1.36)\end{tabular} & 
 \begin{tabular}[c]{@{}c@{}}46.2\\(4.73)\end{tabular} & 
 \begin{tabular}[c]{@{}c@{}}71.62\\(0.46)\end{tabular} & 
 \begin{tabular}[c]{@{}c@{}}175.95\\(0.64)\end{tabular} & 
 \begin{tabular}[c]{@{}c@{}}11.27\\(0.59)\end{tabular} & 
 \begin{tabular}[c]{@{}c@{}}15.58\\(0.23)\end{tabular} & 
 \begin{tabular}[c]{@{}c@{}}35.97\\(0.41)\end{tabular} & 
 \begin{tabular}[c]{@{}c@{}}49.07\\(3.69)\end{tabular} & 
 \begin{tabular}[c]{@{}c@{}}75.54\\(0.74)\end{tabular} & 
 \begin{tabular}[c]{@{}c@{}}184.91\\(1.51)\end{tabular} \\ 
\hline
\multirow{2}{*}{\begin{tabular}[c]{@{}c@{}}\textbf{Machine}\\\textbf{Learning}\end{tabular}} & \multicolumn{1}{c|}{Slices (KB)} & \multicolumn{3}{c|}{5.19 (0.44)} & \multicolumn{3}{c|}{4.28 (0.53)} & \multicolumn{3}{c|}{5.87 (0.39)} & \multicolumn{3}{c|}{5.48 (0.39)} & \multicolumn{3}{c|}{5.65 (0.57)} \\ 
\cline{2-17}
 & \begin{tabular}[c]{@{}c@{}}Average Time (s)\\Std Dev\end{tabular} & 
 \begin{tabular}[c]{@{}c@{}}0.37\\(0.02)\end{tabular} & 
 \begin{tabular}[c]{@{}c@{}}0.35\\(0.02)\end{tabular} & 
 \begin{tabular}[c]{@{}c@{}}0.34\\(0.02)\end{tabular} & 
 \begin{tabular}[c]{@{}c@{}}0.30\\(0.02)\end{tabular} & 
 \begin{tabular}[c]{@{}c@{}}0.30\\(0.02)\end{tabular} & 
 \begin{tabular}[c]{@{}c@{}}0.36\\(0.02)\end{tabular} & 
 \begin{tabular}[c]{@{}c@{}}0.38\\(0.01)\end{tabular} & 
 \begin{tabular}[c]{@{}c@{}}0.43\\(0.01)\end{tabular} & 
\begin{tabular}[c]{@{}c@{}} 0.43\\(0.01)\end{tabular} & 
 \begin{tabular}[c]{@{}c@{}}0.32\\(0.01)\end{tabular} & 
 \begin{tabular}[c]{@{}c@{}}0.35\\(0.01)\end{tabular} & 
 \begin{tabular}[c]{@{}c@{}}0.32\\(0.01)\end{tabular} & 
 \begin{tabular}[c]{@{}c@{}}0.41\\(0.02)\end{tabular} & 
 \begin{tabular}[c]{@{}c@{}}0.43\\(0.03)\end{tabular} & 
 \begin{tabular}[c]{@{}c@{}}0.40\\(0.02)\end{tabular} \\
\hline
\end{tabular}
\end{table*}

\subsubsection{Training}\hfill\\


The dataset is divided into two separate sets, one for training and testing while the second set is for validation purposes. 
The training samples are divided into training and testing subsets. 
We generate approximately 47 million
samples of 128 bytes in size from the training subset of heap data. The dataset is highly imbalanced as we only have 550K positive samples (slices that contain a key) with an imbalance ratio of approximately 1:100. 

We use a stacked classifier approach, consisting of a high precision classifier and a high recall classifier, to predict whether a slice contains a key or not. 
A random forest for the final prediction fed by the probabilities of the high precision and high recall classifiers completes the whole ensemble.
We train the dataset on the Scikit-learn implementation of Random Forest Classifiers, with the only non-default parameter being the number of estimators set as 5 (n\_estimators=5)\cite{scikit-learn}. 
The random forest trained on the actual imbalanced dataset yields our high precision classifier.
The data is oversampled using Synthetic Minority Over-sampling Technique (SMOTE) with the imbalanced learn library\cite{JMLR:v18:16-365}, and another random forest classifier is fit on this data. The high recall classifier is the random forest trained on the oversampled data.



Using the method described, we predict the presence of keys within a 128-byte slice of data. 
Then, the brute force method, described in Section~\ref{subsec:baseline}, extracts the actual keys from the 128-byte window.

\section{Results and Discussion}\label{sec:result-discussion}
While the brute-force method is exact, it exhibits a high runtime-complexity. Thus, when comparing the performance between the machine learning and the brute-force method, we are interested in the trade-off between runtime performance and missing out keys. 

\subsection{Machine Learning}
We compare our Random Forest Classifier based results to a method of key extraction using the optimised brute-force method. As for the machine learning method, we opt for two different methods to analyze results. The first one is the standard way of computing the metrics such as Precision, Recall, F1-Score and Accuracy. The second method is by checking how many encryption keys are retrieved from a corresponding heap. There are overlaps in the slices, and keys can be present in more than one slice. Therefore, the successful keys extraction results from identifying either of the slices where the key is present. \autoref{tab:ml_result} shows the metrics obtained using the machine learning method. \autoref{tab:ml_result_a} shows the four commonly used metrics computed using the scikit-learn library. We see that the high recall classifier can quickly identify most keys, only missing out on one key for the 12-byte and 32-byte length keys while successfully retrieving all keys of length 24 bytes. The only exception is the key length 16, where we miss 54 of the keys, and even then, our worst recall is at 99.78\%. These results show that the random forest performs exceptionally well in identifying data slices that possibly contain keys. 

\subsection{Performance}

We generate a total of 1,500 entirely new data samples that comprise of three different encryption key lengths (16, 24 and 32 bytes) and five different OpenSSH versions, including the recording of the network traffic.
We run the brute-force and the machine learning methods and compare the respective performances.
For this performance comparison, we focus on extracting only Key A and Key C.
Our test-bed is a mini PC that has AMD Ryzen 5500U, 16GB of RAM, 1TB of NVMe SSD and running Debian 11 as the operating system.

As shown in \autoref{tab:performance-result}, the pre-processing step of the brute-force method cleans the heap and reduces the size to around 30\% of the original size. On the other hand, the machine learning method is able to produce slices that hold the potential keys with less than 2\% in size compared to the original size of the heap dump.

The time for the machine learning method includes loading the model, extracting the slices, and using the brute force method to find the key. Nevertheless, it is still significantly faster than the brute-force method. The brute-force method performs poorly whenever the keys are present at the end of the heap dump (e.g. on version V\_7\_9\_P1 and V\_8\_1\_P1). 

\section{Conclusion}\label{sec:conclusion}
While there are other datasets for memory forensics, to the best of our knowledge, our dataset is the first dataset in SSH Key Extraction across different scenarios and different versions and key lengths. Additionally, we also provide the corresponding tool-chain to further create or extend the data-set. We show that while brute-force methods are inherently effective; machine learning algorithms can efficiently produce models with very high recall or precision depending on the scenario. 
The machine learning method has some advantages over the brute-force method as it does not require the coding of any domain knowledge into the system, and does not require the network traffic data.

\section{Acknowledgement}\label{sec:Acknowledgement}
This work has been funded by the 
Bundesministerium für Bildung und Forschung (BMBF, German Federal Ministry of Education and Research) -- project 01IS21063A-C (SmartVMI).

\bibliographystyle{Bibliography/bibliography}
\bibliography{Bibliography/bibliography}
\end{document}